\input ./style/arxiv-general.cfg
\documentclass[aoas,MSNbibl,nameyear,dvips]{arximspdf}
\makeatletter
   \@ifpackageloaded{graphicx}{}{\usepackage{graphicx}}
\makeatother

\doi{10.1214/14-AOAS802}
\volume{9}
\issue{1}
\pubyear{2015}
\firstpage{429}
\lastpage{451}
\docsubty{FLA}

\makeatletter
\newcommand{\eqref}[1]{(\ref{#1})}
\makeatother

\begin{document}
\begin{frontmatter}

\title{A Markov random field-based approach to
characterizing human brain development using spatial--temporal
transcriptome data\thanksref{T1}}
\runtitle{Analysis of human brain development}

\begin{aug}
\author[A]{\fnms{Zhixiang}~\snm{Lin}\thanksref{m1}\ead[label=e1]{zhixiang.lin@yale.edu}},
\author[B]{\fnms{Stephan J.}~\snm{Sanders}\thanksref{m2}\ead[label=e2]{stephan.sanders@ucsf.edu}},
\author[C]{\fnms{Mingfeng}~\snm{Li}\thanksref{m1}\ead[label=e3]{mingfeng.li@yale.edu}},
\author[C]{\fnms{Nenad}~\snm{Sestan}\thanksref{m1}\ead[label=e4]{nenad.sestan@yale.edu}},
\author[B]{\fnms{Matthew W.}~\snm{State}\thanksref{m2}\ead[label=e5]{matthew.state@ucsf.edu}}
\and
\author[F]{\fnms{Hongyu}~\snm{Zhao}\corref{}\thanksref{m1}\ead[label=e6]{hongyu.zhao@yale.edu}}
\runauthor{Z. Lin et al.}
\affiliation{Yale University\thanksmark{m1} and University of
California, San Francisco\thanksmark{m2}}
\address[A]{Z. Lin\\
Interdepartmental Program in Computational\\
\quad Biology and Bioinformatics\\
Yale University\\
New Haven, Connecticut 06511\\
USA\\
\printead{e1}}
\address[B]{S. J. Sanders \\
M. W. State\\
Department of Psychiatry\\
University of California\\
San Francisco, California 94143\\
USA\\
\printead{e2}\\
\phantom{E-mail:\ }\printead*{e5}}
\address[C]{M. Li\\
N. Sestan\\
Department of Neurobiology\\
Kavli Institute for Neuroscience\\
Yale University\\
New Haven, Connecticut 06520\\
USA\\
\printead{e3}\\
\phantom{E-mail:\ }\printead*{e4}}
\address[F]{H. Zhao \\
Department of Biostatistics\\
Yale School of Public Health\\
Yale University\\
New Haven, Connecticut 06520\hspace*{7pt}\\
USA\\
\printead{e6}}
\end{aug}
\thankstext{T1}{Supported in part by NIH GM059507, MH081896, CA154295,
NSF DMS 1106738.}

%
\received{\smonth{10} \syear{2013}}
%
\revised{\smonth{11} \syear{2014}}

\begin{abstract}
Human neurodevelopment is a highly regulated biological process. In
this article, we study the dynamic changes of neurodevelopment through
the analysis of human brain microarray data, sampled from 16 brain
regions in 15 time periods of neurodevelopment. We develop a two-step
inferential procedure to identify expressed and unexpressed genes and
to detect differentially expressed genes between adjacent time periods.
Markov Random Field (MRF) models are used to efficiently utilize the
information embedded in brain region similarity and temporal dependency
in our approach. We develop and implement a Monte Carlo
expectation--maximization (MCEM) algorithm to estimate the model
parameters. Simulation studies suggest that our approach achieves lower
misclassification error and potential gain in power compared with
models not incorporating spatial similarity and temporal dependency.
\end{abstract}

\begin{keyword}
\kwd{Markov Random Field model}
\kwd{spatial and temporal data}
\kwd{neurodevelopment}
\kwd{microarray}
\kwd{Monte Carlo expectation--maximization algorithm}
\kwd{gene expression}
\kwd{differential expression}
\end{keyword}
\end{frontmatter}

\section{Introduction}\label{sec1}

Human neurodevelopment is a dynamic and highly regulated biological
process. Abnormalities in neurodevelopment may lead to psychiatric and
neurological disorders, such as Autism Spectrum Disorders (ASD)
[\citet{geschwind2007autism,walsh2008autism,vsestan2012emerging}]. The
statistical methodology developed in this paper was motivated by our
interest in studying human brain development using a microarray gene
expression data set, which was collected from 1340 tissue samples of 57
developing and adult post-mortem brains (including 39 with both
hemispheres) [\citet{johnson2009functional,kang2011spatio}]. These 57
post-mortem brains spanned from embryonic development to late
adulthood. A 15-period system, demonstrated in Table~\ref{periodsys},
was defined to represent distinct stages of brain development [\citet
{johnson2009functional,kang2011spatio}]. Except for periods 1 and 2,
tissue samples from 16 brain regions were collected from both
hemispheres in each brain, including the cerebellar cortex (CBC),
mediodorsal nucleus\vadjust{\goodbreak} of the thalamus (MD), striatum (STR), amygdala
(AMY), hippocampus (HIP) and 11 areas of the neocortex, including the
orbital prefrontal cortex (OFC), dorsolateral prefrontal cortex (DFC),
ventrolateral prefrontal cortex (VFC), medial prefrontal cortex (MFC),
primary motor cortex (M1C), primary somatosensory cortex (S1C),
posterior inferior parietal cortex (IPC), primary auditory cortex
(A1C), posterior superior temporal cortex (STC), inferior temporal
cortex (ITC) and the primary visual cortex (V1C) [\citet{johnson2009functional,kang2011spatio}]. Details on the brain regions
are described in the supplementary material Section~1 [\citet{lin2015markovs}].

\begin{table}
\tabcolsep=0pt
\tablewidth=240pt
\caption{The 15-period system in \citet{kang2011spatio}. M, postnatal
months; PCW, post-conceptional weeks; Y, postnatal years}
\label{periodsys}
\begin{tabular*}{239pt}{@{\extracolsep{\fill}}lcc@{}}
\hline
\textbf{Period} & \textbf{Description} & \textbf{Age} \\
\hline
\phantom{0}1 & Embryonic & 4~PCW${}\leq{}$Age${}<{}$8~PCW \\
\phantom{0}2 & Early fetal & 8~PCW${}\leq{}$Age${}<{}$10~PCW \\
\phantom{0}3 & Early fetal & 10~PCW${}\leq{}$Age${}<{}$13~PCW \\
\phantom{0}4 & Early mid-fetal & 13~PCW${}\leq{}$Age${}<{}$16~PCW \\
\phantom{0}5 & Early mid-fetal & 16~PCW${}\leq{}$Age${}<{}$19~PCW \\
\phantom{0}6 & Late mid-fetal & 19~PCW${}\leq{}$Age${}<{}$24~PCW \\
\phantom{0}7 & Late fetal & 24~PCW${}\leq{}$Age${}<{}$38~PCW\\
\phantom{0}8 & Neonatal and early infancy & 0~M (birth)${}\leq{}$Age${}<{}$6~M \\
\phantom{0}9 & Late infancy & 6~M${}\leq{}$Age${}<{}$12~M \\
10 & Early childhood & 1~Y${}\leq{}$Age${}<{}$6~Y \\
11 & Middle and late childhood & 6~Y${}\leq{}$Age${}<{}$12~Y \\
12 & Adolescence & 12~Y${}\leq{}$Age${}<{}$20~Y \\
13 & Young adulthood & 20~Y${}\leq{}$Age${}<{}$40~Y \\
14 & Middle adulthood & 40~Y${}\leq{}$Age${}<{}$60~Y \\
15 & Late adulthood & Age${}\geq{}$60~Y \\
\hline
\end{tabular*}
\end{table}

The goal of our analysis is to characterize human neurodevelopment\break
through the dynamics of gene expression, such as the identification of
expressed and unexpressed genes, and differentially expressed (DE)
genes over time in each brain region. The unique challenge presented
for statistical analysis of this data set is the appropriate modeling
and analysis of the spatial--temporal structure. For gene expression
data with only temporal structure (e.g., time course gene
expression data), various methods have been proposed to model the
temporal dependency to better identify DE genes. However, as far as we
know, none of the existing methods utilizes the information embedded in
the spatial similarity between brain regions, as indicated by the high
correlation in gene expression levels between brain regions in the same
period [supplementary material Section~2, \citet{lin2015markovs} and
\citet{kang2011spatio}]. For time course gene expression data, the
existing methods can be classified into two broad categories: (1)
methods that identify DE genes between multiple\vadjust{\goodbreak} biological conditions
[\citet{storey2005significance,hong2006functional}, Tai and Speed (\citeyear{tai2006multivariate}), \citet{yuan2006hidden}]; and (2) methods that identify DE genes over time in
one biological condition [\citet{storey2005significance,tai2006multivariate,wu2007statistical,liu2009identifying}].
Statistical models that have been proposed to incorporate the temporal
structure include Hidden Markov Models [\citet{yuan2006hidden,wu2007statistical}], functional models using basis
function expansions [\citet
{storey2005significance,hong2006functional,wu2007statistical}],
function principal component analysis [\citet{liu2009identifying}] and
multivariate empirical Bayes models [\citet{tai2006multivariate}].

To efficiently capitalize on brain region similarity and temporal
dependency, we propose a two-step Markov Random Field (MRF)-based
approach to answer the following two biological questions: 1. Which
genes are expressed/unexpressed in each period and in each brain
region? 2. Which genes are differentially expressed over time in each
brain region? We note that MRF models have been used to model
dependency in genomics data, such as neighboring genes defined by
biological pathways [\citet{li2010network,chen2011incorporating}, \citeauthor{wei2007markov}
(\citeyear{wei2007markov,wei2007statistical})]
and marker dependencies defined by linkage disequilibrium [\citet
{li2010hidden}]. Across all the brain regions and time periods, the
histogram of the observed gene expression levels has a bimodal
distribution, where the two components likely represent expressed and
unexpressed genes [supplementary material Section~4, \citet
{lin2015markovs} and \citet{kang2011spatio}]. In this paper, we first
use a Gaussian mixture model-based approach to identify the unexpressed
and expressed genes. The model fit and the robustness of the Gaussian
mixture model are discussed in the supplementary material
Section~4
[\citet{lin2015markovs}]. We note that an ``unexpressed'' gene does not
necessarily suggest that there is no mRNA molecules of that gene in the
cell, but rather the gene's expression level is very low and the
observed variation in the expression values may be mostly due to noise
in the microarray experiment. In the second step, our methodology
utilizes the local false discovery rate (f.d.r.) framework [\citet
{efron2004large}] to identify DE genes between adjacent time periods.
We propose an efficient Monte Carlo expectation--maximization (MCEM)
algorithm [\citet{wei1990monte}] to estimate the model parameters and a
Gibbs sampler to estimate the posterior probabilities.

The key feature of our approach is to simultaneously consider spatial
similarity and temporal dependency of gene expression levels to better
extract biologically meaningful results from the data. We introduce the
MRF model in Section~\ref{sec2} and present the Monte Carlo
expectation--maximization (MCEM) algorithm for statistical inference in
Section~\ref{sec3}. We also present the posterior probability estimation and the
FDR controlling procedure in Section~\ref{sec3}. In Section~\ref{sec4} we apply our
method to analyze the human brain microarray data reported in \citet{kang2011spatio}. Results from simulation studies are summarized in
Section~\ref{sec5}. We conclude the paper with a brief discussion in\vadjust{\goodbreak} Section~\ref{sec6}.


\section{Statistical models and methods}\label{sec2}
\subsection{Biological question 1: Identify expressed and unexpressed genes}\label{sec21}
\subsubsection{Gaussian mixture model for microarray data}\label{sec211}
In our human brain microarray data, expression levels were measured for
$G=17{,}568$ genes on the Affymetrix GeneChip Human Exon 1.0 ST Array
platform. For quality control, RMA background correction, quantile
normalization, mean probe set summarization and
$\log_{2}$-transformation were performed [\citet{kang2011spatio}].
Details for the quality control procedures are described in the
supplementary material Section~3 [\citet{lin2015markovs}]. The number
of brains that were collected varies across time periods and for some
brains, tissue samples are missing for certain brain regions. So the
number of samples varies among brain regions and time periods. We
treated samples from the same brain region and time period as
biological replicates. Periods 1 and 2 correspond to embryonic and
early fetal development, when most of the 16 brain regions sampled in
future periods have not differentiated (i.e., most of the 16 brain
regions are missing data in periods 1 and~2). Therefore, samples in
periods 1 and 2 are excluded in our analysis. In total, we consider
$B=16$ brain regions sampled in $T=13$ periods of brain development.
Let $n_{bt}$ denote the number of replicates for brain region $b$ in
period $t$, $\mathbf{N}_{b}=(n_{b1},\ldots,n_{bt},\ldots,n_{bT})'$ is the
column vector for the number of replicates for brain region $b$, and
$\mathbf{N}=(\mathbf{N}_{1},\ldots,\mathbf{N}_{b},\ldots,\mathbf{N}_{B})$ is
the matrix summarizing the number of replicates across brain regions
and periods. The entries in $\mathbf{N}$ range from 1 to 16 and the
median is 5. Let $y_{bgtk}$ denote the observed gene expression value
for gene $g$ in the $k$th replicate of samples in brain region $b$ and
period $t$, and let $\mathbf{y}_{bgt}=(y_{bgt1},\ldots,y_{bgtn_{bt}})$
denote the expression values for all the replicates. We assume that
$y_{bgtk}$,  for $k=1,\ldots,n_{bt}$, follows the same normal
distribution with mean $\mu_{bgt}$ and standard deviation $\sigma_{0}^{2}$:
\[
y_{bgtk}\sim\mathcal{N} \bigl(\mu_{bgt},\sigma_{0}^{2}
\bigr).
\]
Let $x_{bgt}$ be the binary latent state representing whether gene $g$
is expressed in brain region $b$ and period $t$, that is, $x_{bgt}=1$ if
the gene is expressed and $0$ otherwise. Conditioning on $x_{bgt}$, we
assume that $\mu_{bgt}$ follows a Gaussian distribution:
\begin{eqnarray*}
\mu_{bgt}\vert x_{bgt} &=& 0 \sim\mathcal{N} \bigl(
\mu_{1b},\sigma_{1b}^{2} \bigr),
\\
\mu_{bgt}\vert x_{bgt} &=& 1 \sim\mathcal{N} \bigl(
\mu_{2b},\sigma_{2b}^{2} \bigr).
\end{eqnarray*}
Marginally, $\mu_{bgt}$ follows a Gaussian mixture distribution. We
assume that the mean and the variance for the mixture components are
brain region specific. Denote by $\bolds{\mu}_{1},\bolds{\mu}_{2},
\bolds{\sigma}_{1},\bolds{\sigma}_{2}$ the vectors of parameters for all brain regions.
It is easy to see that the distribution of $y_{bgtk}$ conditioning on
$x_{bgt}$ has the following form:
\begin{eqnarray*}
y_{bgtk}\vert x_{bgt}&=&0 \sim\mathcal{N} \bigl(
\mu_{1b},\sigma_{1b}^{2}+\sigma ^{2}_0
\bigr),
\\
y_{bgtk}\vert x_{bgt} &=& 1 \sim \mathcal{N} \bigl(
\mu_{2b},\sigma_{2b}^{2}+\sigma ^{2}_0
\bigr).
\end{eqnarray*}
Given the latent state array $\mathbf{X}$, conditional independence is assumed:
\[
f(\mathbf{Y}\vert \mathbf{X})=\prod_{b=1}^{B}
\prod_{g=1}^{G} \prod
_{t=1}^{T} f(\mathbf{y}_{bgt}\vert
x_{bgt}),
\]
where
\[
f(\mathbf{y}_{bgt}\vert x_{bgt})=\prod
_{k=1}^{n_{bt}}f(y_{bgtk}\vert x_{bgt}).
\]

\subsubsection{A MRF model for $p(\mathbf{X})$}\label{sec212}
One key component in the above model and the inferential objective is
the latent state array $\mathbf{X}$, which is unknown to us. Now we
discuss how to specify the prior on $\mathbf{X}$, denoted by
$p(\mathbf{X})$, through a MRF model that takes into account both temporal
dependency and spatial similarity. For each gene $g$, we construct an
undirected graph $G_g=\{V_g,E_g\}$, where $V_g=\{x_{bgt}\dvtx  b=1,\ldots,B,
t=1,\ldots,T\}$ is the set of nodes and $E_g$ is the set of edges. $E_g$
can be divided into two subsets, $E_{g1}$ and $E_{g2}$, where $E_{g1}=\{
(x_{bgt},x_{b'gt'})\dvtx  b\neq b' \mbox{ and } t=t' \}$ and $E_{g2}=\{
(x_{bgt},x_{b'gt'})\dvtx b=b'\mbox{ and }|t-t'|=1\}$. $E_{g1}$ contains the
edges capturing spatial similarity between brain regions and $E_{g2}$
contains the edges capturing temporal dependency between adjacent
periods. For the joint distribution of $p(\mathbf{X})$, we construct a
pairwise interaction MRF model [\citet{besag1986statistical}] with the
following form:
%
\begin{eqnarray}
p(\mathbf{X}\vert \bolds{\Phi} ) & \propto & \prod_{g=1}^{G}
\exp \biggl\{\gamma_{0}\sum_{V_{g}}
I_{0}(x_{bgt})+ \gamma_{1}\sum
_{V_{g}} I_{1}(x_{bgt})
\nonumber
\\
\label{jointunexp} &&\qquad\hspace*{14pt} {}+\beta_{1}\sum
_{E_{g1}} \bigl[ I_0(x_{bgt})I_0(x_{b'gt'})+I_1(x_{bgt})I_1(x_{b'gt'})
\bigr]
\\
&&\qquad\hspace*{14pt} {}+\beta_{2}\sum_{E_{g2}}
\bigl[ I_0(x_{bgt})I_0(x_{b'gt'})+I_1(x_{bgt})I_1(x_{b'gt'})
\bigr] \biggr\},
\nonumber
\end{eqnarray}
where $I_{0}(\cdot)$ and $I_{1}(\cdot)$ are the indicator functions.
Letting $\gamma=\gamma_1-\gamma_0$, the conditional probability can be
derived (see \hyperref[app]{Appendix} for the details of derivation):
%
\begin{equation}
\label{cp} p(x_{bgt}\vert \mathbf{X}/x_{bgt};\bolds{\Phi})=
\frac{\exp\{x_{bgt}F(x_{bgt},\bolds{\Phi})\}}{1+\exp\{F(x_{bgt},\bolds{\Phi})\}},
\end{equation}
where
\begin{eqnarray*}
F(x_{bgt},\bolds{\Phi}) &=& \gamma + \beta_{1}\sum
_{b'\neq
b}(2x_{b'gt}-1)
\\
&&{}+\beta_{2} \bigl\{I_{t \neq1}[2x_{bg(t-1)}-1]+I_{t \neq
T}[2x_{bg(t+1)}-1]
\bigr\},
\end{eqnarray*}
where ``$/$'' means other than; $\bolds{\Phi}=(\gamma,\beta_{1},\beta_{2})$
and $\gamma,\beta_{1},\beta_{2}\in\mathbf{R}$; $\beta_{1}$ is the
parameter capturing the spatial similarity and $\beta_{2}$ is the
parameter capturing the temporal dependency.
\subsection{Biological question 2: Identify DE genes over time}\label{sec22}

\subsubsection{A latent state model for DE}\label{sec221}
For DE analysis, we first transform the observed data into an array
where the entries are then used in the follow-up analysis. This is
accomplished by performing $t$-tests between adjacent periods and
transforming the $t$-statistics into $z$-scores. Let $\mathbf{y}_{bg(t-1)}$ and $\mathbf{y}_{bgt}$ denote the vectors of expression
values for gene $g$ in region $b$ and in periods $t-1$ and $t$,
respectively. The two-sample $t$-statistic is obtained by
\[
t_{bg(t-1)}=\frac{\bar{\mathbf{y}}_{bgt}-\bar{\mathbf{y}}_{bg(t-1)}}{s},
\]
where $s$ is an estimate of the standard error for $\bar{\mathbf
{y}}_{bgt}-\bar{\mathbf{y}}_{bg(t-1)}$. The test statistic
$t_{bg(t-1)}$ is then transformed into $z_{bg(t-1)}$:
\[
z_{bg(t-1)}=\Phi^{-1} \bigl(F_{n_{bt}+n_{b(t-1)}-2}(t_{bg(t-1)})
\bigr),
\]
where $n_{b(t-1)}$ and $n_{bt}$ are the numbers of replicates in
$\mathbf{y}_{bg(t-1)}$ and $\mathbf{y}_{bgt}$; $\Phi$ and
$F_{n_{bt}+n_{b(t-1)}-2}$ are the c.d.f.s for standard normal and $t$
distribution with ${n_{bt}+n_{b(t-1)}-2}$ degrees of freedom. As a
result, the gene expression data are represented by a $B\times G \times
(T-1)$ $z$-score array $\mathbf  Z$. The entry $z_{bgt}$ represents the
evidence of DE between periods $t$ and $t+1$ for gene $g$ in brain
region $b$. Some entries in the array are not assigned values because
of the presence of unexpressed genes. The variations in the expression
values of unexpressed genes may be mostly caused by noise in the
microarray experiments and we do not want to include that noise in
identifying DE genes; the transitions from unexpressed to expressed and
vice versa are already captured in biological question 1. Therefore, no
$t$-test is performed if the gene is unexpressed in at least one of the
adjacent periods. Let $s_{bgt}$ denote the binary latent state
representing whether gene $g$ is differentially expressed in brain
region $b$ between periods $t$ and $t+1$, which is the objective of our
inference. Let $\mathbf{S}$ be the latent state array of dimensions
$B\times G \times(T-1)$. Conditioning on $s_{bgt}$, we assume that
$z_{bgt}$ follows a mixture distribution:
\[
f(z_{bgt}\vert s_{bgt})=(1-s_{bgt})f_{0}(z_{bgt})+s_{bgt}f_{1}(z_{bgt}),
\]
where $f_{0}(z)$ is the null density and $f_{1}(z)$ is the nonnull
density. We assume that the null density follows a standard normal
$\mathcal{N}(0,1)$ distribution. We adopt the nonparametric empirical
Bayesian framework for DE [\citet{efron2004large}] by fitting the
nonnull density with a natural spline using the R package $\mathit{locfdr}$.
Given $\mathbf{S}$, conditional independence is assumed:
\[
f(\mathbf{Z}\vert \mathbf{S})=\prod_{b=1}^{B}
\prod_{g=1}^{G} \prod
_{t=1}^{T-1} f(z_{bgt}\vert s_{bgt}).
\]

\subsubsection{A MRF model for $p(\mathbf{S})$}\label{sec222}
Next, we present a MRF model for the prior distribution $p(\mathbf{S})$, taking into account both temporal
dependency and spatial similarity. We separate the 16 brain regions
into two groups: 11 neocortex regions, represented by $\mathbf{B}_{c}$, and 5
nonneocortex regions, represented by $\mathbf{B}_{n}$. The joint probability
is similar to \eqref{jointunexp}, except that different spatial
parameters are assumed for the two groups. The conditional probability
can be calculated and has the following form:
%
\begin{equation}
\label{cpDE} p(s_{bgt}\vert \mathbf{S}/s_{bgt};\bolds{
\Phi}_{\mathrm{DE}})= \frac{\exp\{
s_{bgt}F_{\mathrm{DE}}(s_{bgt},\bolds{\Phi}_{\mathrm{DE}})\}}{1+\exp\{F_{\mathrm{DE}}(s_{bgt},\bolds{\Phi
}_{\mathrm{DE}})\}},
\end{equation}
if $b\in\mathbf{B}_{c}$,
\begin{eqnarray*}
F_{\mathrm{DE}}(s_{bgt},\bolds{\Phi}_{\mathrm{DE}})&=&
\gamma_{\mathrm{DE}} + \beta_{\mathrm{cc}}\sum_{b'\in\mathbf{B_c}/b}(2s_{b'gt}-1)
+ \beta_{\mathrm{cn}}\sum_{b'\in\mathbf{B}_{n}}(2s_{b'gt}-1)
\\
&&{}+\beta_{t} \bigl\{I_{t \neq1}[2s_{bg(t-1)}-1]+I_{t \neq
T}[2s_{bg(t+1)}-1]
\bigr\},
\end{eqnarray*}
else if $b\in\mathbf{B}_{n}$,
\begin{eqnarray*}
F_{\mathrm{DE}}(s_{bgt},\bolds{\Phi}_{\mathrm{DE}}) &=&
\gamma_{\mathrm{DE}} + \beta_{\mathrm{nn}}\sum_{b'\in\mathbf{B}_{n/b}}(2s_{b'gt}-1)
+ \beta_{\mathrm{nc}}\sum_{b'\in\mathbf{B}_c}(2s_{b'gt}-1)
\\
&&{}+\beta_{t} \bigl\{I_{t \neq1}[2s_{bg(t-1)}-1]+I_{t \neq
T}[2s_{bg(t+1)}-1]
\bigr\},
\end{eqnarray*}
where $\bolds{\Phi}_{\mathrm{DE}}=(\beta_{\mathrm{cc}},\beta_{\mathrm{nn}},\beta_{\mathrm{cn}},\beta_{\mathrm{nc}})$,
$\beta_{\mathrm{cc}}$ is the between neocortex coefficient, $\beta_{\mathrm{nn}}$ is the
between nonneocortex coefficient, $\beta_{\mathrm{cn}}$ is the neocortex to
nonneocortex coefficient, and $\beta_{\mathrm{nc}}$ is the nonneocortex to
neocortex coefficient. For symmetry, we assume that $\beta_{\mathrm{cn}}=\beta
_{\mathrm{nc}}$. In the MRF model in Section~\ref{sec212}, we did not separate the
brain regions into two groups because the latent states for all brain
regions were quite similar, which will be shown in Section~\ref{sec4}.

\section{Parameter and posterior probability estimation}\label{sec3}
\subsection{Parameter estimation for biological question 1: Identify
expressed and unexpressed genes}\label{sec31}
In the model, the MRF parameters $\bolds{\Phi}=(\gamma,\beta_{1},\beta
_{2})$ and the Gaussian mixture model parameters $\bolds{\Theta}=(\bolds{\mu}_{1},\bolds{\sigma}_{1},
\bolds{\mu}_{2},\bolds{\sigma}_{2})$ need to be
estimated. Given the latent state $\mathbf{X}$, both $\bolds{\Phi}$ and
$\bolds{\Theta}$ can be estimated by the maximum likelihood estimates
(MLE). However, the latent state is unobserved and needs to be
estimated as well. Although the expectation--maximization (EM) algorithm
is generally implemented for missing data estimation, it is not
applicable to our model as the expectation term is not tractable.
Therefore, we propose the following Monte Carlo EM Algorithm [\citet
{wei1990monte}] to estimate $\bolds{\Phi}$ and $\bolds{\Theta}$:
\begin{longlist}[4.]
\item[1.] Estimate $\sigma_{0}$ by the unbiased estimator:
\[
\hat{\sigma}_{0}^{2}=\frac{1}{G\times\sum_{b=1}^{B}\sum_{t=1}^{T}(n_{bt}-1)} \sum
_{g=1}^{G}\sum_{b=1}^{B}
\sum_{t=1}^{T}\sum
_{k=1}^{n_{bt}}(y_{bgtk}-\bar{y}_{bgt})^2.
\]
\item[2.] Obtain the initial estimates $\hat{\mathbf{X}}$ and
$\hat{\bolds{\Theta}}$ by the simple Gaussian mixture model, without considering
spatial and temporal dependency.
\item[3.] Because there is no explicit MLE for $\bolds{\Phi}$, an initial
estimate $\hat{\bolds{\Phi}}$ is chosen which maximizes the following
pseudolikelihood function $l(\hat{\mathbf{X}};\bolds{\Phi})$ [\citet{besag1974spatial}]:
\[
l(\hat{\mathbf{X}};\bolds{\Phi})=\prod_{b=1}^{B}
\prod_{g=1}^{G}\prod
_{t=1}^{T} p(\hat{x}_{bgt}\vert \hat{
\mathbf{X}}/\hat{x}_{bgt};\bolds{\Phi}),
\]
where $p(\hat{x}_{bgt}\vert \hat{\mathbf{X}}/\hat{x}_{bgt};\bolds{\Phi})$
is as defined in \eqref{cp}.
\item[4.] Let $\bolds{\Psi}=(\bolds{\Phi},\bolds{\Theta})$. The expected complete
data log-likelihood in the EM algorithm is approximated by the Monte
Carlo sum [\citet{wei1990monte}]:
%
\begin{equation}
\label{qfunc} Q_{m} \bigl(\bolds{\Psi}\vert \hat{\bolds{\Psi}}^{(r)} \bigr)= \frac{1}{m}\sum_{l=1}^{m}
\ln f \bigl( \mathbf{Y},\mathbf{X}_{l}^{(r)}\vert \bolds{\Psi}
\bigr),
\end{equation}
where $\mathbf{X}_{1}^{(r)},\ldots,\mathbf{X}_{m}^{(r)}$ are obtained by Gibbs
sampling. From $\mathbf{X}_{l}^{(r)}$ to $\mathbf{X}_{(l+1)}^{(r)}$, all
entries in $\mathbf{X}_{l}^{(r)}$ are updated, and they are updated
sequentially by
%
\begin{equation}
\label{postprob} p \bigl(x_{bgt}\vert \mathbf{Y},\mathbf{X}/x_{bgt};
\hat{\bolds{\Psi}}^{(r)} \bigr)\propto p \bigl(x_{bgt}\vert
\mathbf{X}/x_{bgt};\hat{\bolds{\Phi}}^{(r)} \bigr) f \bigl(
\mathbf{y}_{bgt}\vert x_{bgt};\hat{\bolds{\Theta}}^{(r)}
\bigr).
\end{equation}

\item[5.] Update $\bolds{\Psi}$ by $\hat{\bolds{\Psi}}^{(r+1)}$, which maximizes
\eqref{qfunc}:
\[
\hat{\bolds{\Psi}}^{(r+1)}=\mathop{\arg\max}_{\bolds{\Psi}}Q_{m}
\bigl(\bolds{\Psi}\vert \hat {\bolds{\Psi}}^{(r)} \bigr).
\]
Same as in step 3, we replace the likelihood by the pseudolikelihood
function in $Q_{m}(\bolds{\Psi}\vert \hat{\bolds{\Psi}}^{(r)})$. The terms
that contain $\bolds{\Phi}$ and $\bolds{\Theta}$ are separable, therefore,
they can be optimized separately.
%
\item[6.] Repeat steps 4 and 5 until convergence.
\end{longlist}
%
\subsection{Parameter estimation for biological question 2: Identify DE
genes over time}\label{sec32}
In the model, only the parameters $\bolds{\Phi}$ in the MRF prior need to
be updated iteratively. The algorithm shares some similarity with that
in the previous section:
\begin{longlist}[6.]
\item[1.] Pool the $z$-scores in $\mathbf{Z}$ and estimate $f_{1}$ by the
$\mathit{locfdr}$ procedure.
\item[2.] Obtain an initial estimate $\hat{\mathbf{S}}$ by the simple
mixture model, without considering spatial and temporal dependency.
\item[3.] Obtain an initial estimate $\hat{\bolds{\Phi}}_{\mathrm{DE}}$, which
maximizes the pseudolikelihood function:
\[
l(\hat{\mathbf{S}};\bolds{\Phi}_{\mathrm{DE}})=\prod
_{b=1}^{B}\prod_{g=1}^{G}
\prod_{t=1}^{T-1} p(\hat{s}_{bgt}
\vert \hat{\mathbf{S}}/\hat{s}_{bgt};\bolds{\Phi}_{\mathrm{DE}}),
\]
where $p(\hat{s}_{bgt}\vert \hat{\mathbf{S}}/\hat{s}_{bgt};\bolds{\Phi}_{\mathrm{DE}})$ is as defined in \eqref{cpDE}.
\item[4.] Approximate the expected complete data log-likelihood by the
Monte Carlo sum:
%
\begin{equation}
\label{qfuncDE} Q_{m} \bigl(\bolds{\Phi}_{\mathrm{DE}}\vert \hat{
\bolds{ \Phi}}_{\mathrm{DE}}^{(r)} \bigr)=\frac
{1}{m}\sum
_{l=1}^{m}\ln f \bigl(\mathbf{Z},
\mathbf{S}_{l}^{(r)} \vert \bolds{\Phi}_{\mathrm{DE}} \bigr),
\end{equation}
where $\mathbf{S}_{1}^{(r)},\ldots,\mathbf{S}_{m}^{(r)}$ are obtained by Gibbs
sampling. From $\mathbf{S}_{l}^{(r)}$ to $\mathbf{S}_{(l+1)}^{(r)}$, all
entries in $\mathbf{S}_{l}^{(r)}$ are updated, and they are updated
sequentially by
%
\begin{equation}
\label{postDE} p \bigl(s_{bgt}\vert \mathbf{Z},\mathbf{S}/s_{bgt};
\hat{\bolds{\Phi}}_{\mathrm{DE}}^{(r)} \bigr)\propto p
\bigl(s_{bgt}\vert \mathbf{S}/s_{bgt};\hat{\bolds{
\Phi}}_{\mathrm{DE}}^{(r)} \bigr) f(z_{bgt}\vert
s_{bgt}).
\end{equation}

\item[5.] Update $\bolds{\Phi}_{\mathrm{DE}}$ by $\hat{\bolds{\Phi}}_{\mathrm{DE}}^{(r+1)}$, which maximizes \eqref{qfuncDE}.
\item[6.] Repeat steps 4 and 5 until convergence.
\end{longlist}
%
\subsection{Posterior probability estimation and FDR controlling procedure}\label{sec33}
To acquire an estimate of the posterior probability, we implement a
separate Gibbs sampler and keep the model parameters fixed at the
estimated values by the MCEM algorithm. The latent states in biological
questions 1 and 2 are updated sequentially according to \eqref
{postprob} and \eqref{postDE}.

For the inference of expressed/unexpressed genes, we use 0.5 as the
cutoff for the posterior probability. For the inference of DE genes, we
adapt the posterior probability-based definition of FDR [\citet{newton2001differential,li2010hidden}]. The posterior local f.d.r.
$q_{bgt}=p(s_{bgt}=0\vert \mathbf{Z})$ is estimated by the Gibbs sampler.
Let $q_{(s)}$ be the sorted values of $q_{bgt}$ in ascending order.
Find $k=\max\{t\dvtx \frac{1}{t}\sum_{s=1}^{t}q_{(s)}\leq\alpha\}$ and
reject all the null hypotheses $H_{(s)}, \mbox{ for } s=1,\ldots,k$. In
the analysis of human brain gene expression data, we chose $\alpha=0.05$.
\section{Application to the human brain microarray data}\label{sec4}
\subsection{Identify expressed and unexpressed genes}\label{sec41}
We first applied the MRF model to infer whether a gene is expressed or
not in a certain brain region and time period. In the parameter
estimation, we first ran 20 iterations of MCEM by a Gibbs sampler with
$500/1500$ (1500 iterations in total and 500 as burn-in), then 20
iterations with $1000/6000$ and, finally, 20 iterations with
$1000/10{,}000$. We gradually increased the number of iterations in the
Gibbs sampler to make the estimate of the parameters more stable. The
posterior probability was then estimated by a Gibbs sampler with
$10{,}000$
iterations and 1000 as burn-in. A diagnosis for the number of
iterations is presented in the supplementary material Section~5
[\citet{lin2015markovs}].
%
\begin{table}[b]
\tablewidth=220pt
\caption{The estimated parameters for the Gaussian mixture model}
\label{GMMparaT}
\begin{tabular*}{220pt}{@{\extracolsep{\fill}}lcccc@{}}
\hline
\textbf{Region} & $\bolds{\mu_1}$ & $\bolds{\mu_2}$ & $\bolds{\sigma_1}$ & $\bolds{\sigma_2}$ \\
\hline
MFC & 4.58 & 7.82 & 0.59 & 1.57 \\
OFC & 4.57 & 7.83 & 0.59 & 1.58 \\
VFC & 4.56 & 7.84 & 0.58 & 1.59 \\
DFC & 4.58 & 7.83 & 0.58 & 1.58 \\
STC & 4.62 & 7.8\phantom{0} & 0.58 & 1.56 \\
ITC & 4.61 & 7.81 & 0.58 & 1.57 \\
A1C & 4.6\phantom{0} & 7.82 & 0.58 & 1.57 \\
IPC & 4.61 & 7.81 & 0.58 & 1.57 \\
S1C & 4.61 & 7.82 & 0.58 & 1.58 \\
M1C & 4.60 & 7.82 & 0.58 & 1.58 \\
V1C & 4.63 & 7.78 & 0.59 & 1.55 \\
AMY & 4.65 & 7.76 & 0.6\phantom{0} & 1.52 \\
HIP & 4.64 & 7.77 & 0.61 & 1.54 \\
STR & 4.65 & 7.78 & 0.62 & 1.55 \\
MD & 4.62 & 7.81 & 0.63 & 1.59 \\
CBC & 4.61 & 7.76 & 0.65 & 1.58 \\
\hline
\end{tabular*}
\end{table}

\begin{figure}

\includegraphics{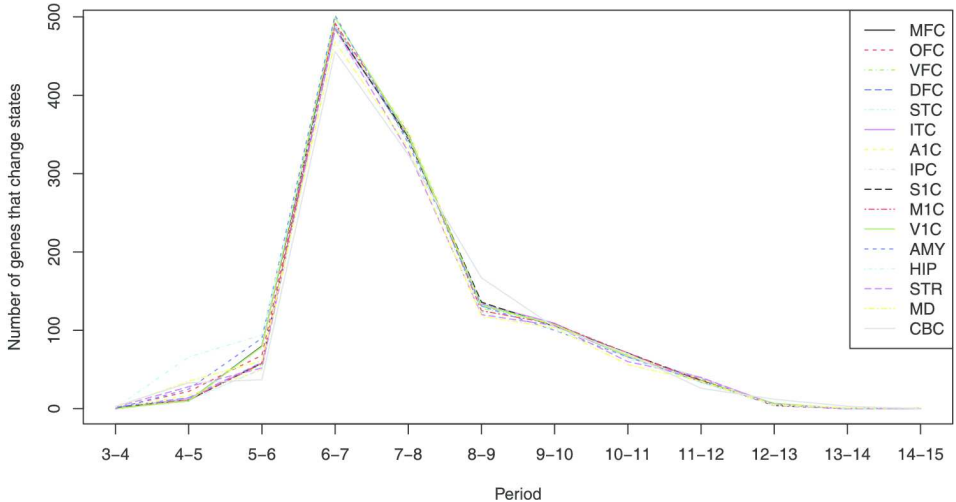}

\caption{The number of genes that changed from expressed to unexpressed
and vice versa in adjacent periods. Each line represents a brain region.}
\label{gmmdatas}
\end{figure}

The estimated parameters for the Gaussian mixture model are shown in
Table~\ref{GMMparaT}. The estimated parameters for the MRF prior were
$\gamma=0.30$, $\beta_{1}=0.22$ and $\beta_{2}=6.44$. The large
coefficient in $\beta_{2}$ indicates strong temporal dependency.
Compared with the total number of genes ($17{,}568$), only a small number
of genes changed their latent states between adjacent periods (Figure~\ref{gmmdatas}). The table for the numbers are presented in
supplementary material Section~8 [\citet{lin2015markovs}]. For all
brain regions, a general trend can be observed: the number of genes
that changed their latent states first increased, peaked in periods 6
to 7, the number in periods 7 to 8 was also large, then gradually
decreased, starting from periods 12 to 13, fewer than 15 genes changed
their latent states. Period 8 corresponds to birth to 6 postnatal
months. The observation that the changes in gene expression peaked from
periods 6 to 8 suggests that robust changes in gene expression occurred
close to birth.

Moreover, we observed that the latent states for the same gene in all
brain regions tended to agree with each other. These are summarized in
Table~\ref{bragree}, where we considered all genes by time
combinations, that is, $G \times T = 17{,}568 \times13 = 228{,}384$, and
counted the number of genes that were expressed in a given number of
brain regions. Although the MRF prior encourages the agreement of
latent states, the observation is unlikely driven by the model, as we
observed a similar trend when the spatial coefficient $\beta_{1}$ was
fixed to be 0 (supplementary material Section~8 [\citet{lin2015markovs}]).

%
\begin{table}
\tablewidth=150pt
\caption{Summary of the latent states by pooling brain regions. ``0''
represents the total count of genes that were unexpressed in all brain
regions and ``16'' represents the total count of genes that were
expressed in all brain regions}
\label{bragree}
\begin{tabular*}{150pt}{@{\extracolsep{\fill}}lc@{}}
\hline
\phantom{0}\textbf{0} & \phantom{0}89{,}347 \\
\phantom{0}\textbf{1} & \phantom{00,}2560 \\
\phantom{0}\textbf{2} & \phantom{000,}541 \\
\phantom{0}\textbf{3} & \phantom{000,}218 \\
\phantom{0}\textbf{4} & \phantom{0000,}95 \\
\phantom{0}\textbf{5} & \phantom{0000,}62 \\
\phantom{0}\textbf{6} & \phantom{0000,}31 \\
\phantom{0}\textbf{7} & \phantom{0000,}52 \\
\phantom{0}\textbf{8} & \phantom{0000,}31 \\
\phantom{0}\textbf{9} & \phantom{0000,}26 \\
\textbf{10} & \phantom{0000,}19 \\
\textbf{11} & \phantom{0000,}46 \\
\textbf{12} & \phantom{0000,}42 \\
\textbf{13} & \phantom{0000,}94 \\
\textbf{14} & \phantom{0000,}99 \\
\textbf{15} & \phantom{000,}297 \\
\textbf{16} & 134{,}824 \\
\hline
\end{tabular*}
\end{table}

Genes that changed states over time may be of biological interest for
the study of brain development. We conducted Gene Ontology (GO)
enrichment analysis using DAVID, which takes a list of genes as input
and outputs the enriched Gene Ontology (GO) terms [\citet{da2008systematic,sherman2009bioinformatics}]. A GO term represents the
functional annotation of a list of genes and may belong to any of the
following three categories: (a) genes that participate in the same
biological process, (b)~genes that have the same molecular function,
and (c) genes that are located in the same cellular component. Only GO
terms in categories (a) and (b) were included in our analysis, as genes
located in the same cellular component do not necessarily share similar
functions. We observed enrichment of GO terms only from periods 6 to~7
(0.05 threshold for Bonferroni-adjusted $p$-value). From periods 6 to
7, genes that switched from expressed to unexpressed in all brain
regions were enriched for ``DNA binding'' (Bonferroni adjusted
$p$-value${}=1.6\times 10^{-9}$), ``regulation of transcription,
DNA-dependent'' (Bonferroni adjusted $p$-value${}=2.5\times10^{-4}$) and
``zinc ion binding'' (Bonferroni adjusted $p$-value${}=9.5\times10^{-5}$);
there were no enriched GO terms for genes that switched from
unexpressed to expressed. The enrichment of transcription regulation
and DNA binding proteins (including zinc-finger proteins coordinated by
the binding of zinc ions) is consistent with our previous observation
that robust changes in transcription occurred close to birth. Changes
in transcriptional regulation may also lead to the peak of
differentially expressed genes (see Section~\ref{sec42}). Details for the GO
enrichment analysis are presented in the supplementary material
Section~6 [\citet{lin2015markovs}].

\subsection{Identify DE genes over time}\label{sec42}
After excluding genes that were unexpressed in all brain regions and
all periods, 11,370 genes remained. We then applied the MRF model to
identify DE genes between adjacent periods. The settings for the MCEM
algorithm and the Gibbs sampler were the same as that in the previous section.

The estimated MRF parameters were $\gamma_{\mathrm{DE}}=-0.10$, $\beta
_{\mathrm{cc}}=0.32$, $\beta_{\mathrm{nn}}=0.53$, $\beta_{\mathrm{cn}}=0.06$, and $\beta
_{t}=0.15$. The temporal coefficient $\beta_{t}$ was much smaller
compared with that in the previous section (where $\beta_{2}=6.44$),
which suggests lower temporal dependency. The neocortex to
nonneocortex coefficient $\beta_{\mathrm{cn}}$ was much smaller than the
neocortex to neocortex coefficient $\beta_{\mathrm{cc}}$ and the nonneocortex
to nonneocortex coefficient $\beta_{\mathrm{nn}}$, which indicates the group
difference between neocortex and nonneocortex regions.

%
\begin{figure}

\includegraphics{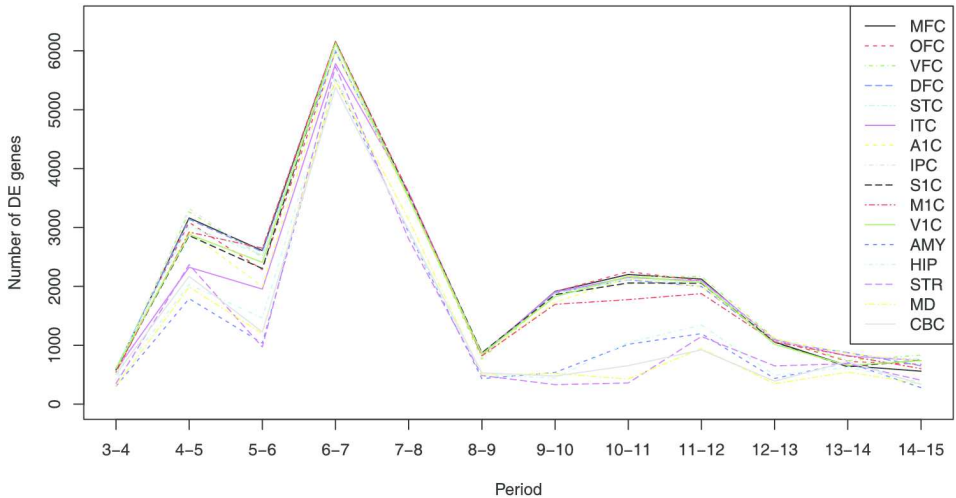}

\caption{The number of DE genes identified in each time window of
adjacent periods. Each line represents a brain region.}
\label{DEdataS}
\end{figure}

When no spatial and temporal dependency is assumed, the model reduces
to a simple empirical Bayesian (EB) model. Based on the posterior FDR
control procedure described in Section~\ref{sec3}, the thresholds in the MRF and
EB models were $0.26$ and $0.12$, respectively. The numbers of genes
identified as DE in the two models were $356{,}207$ (MRF) and $77{,}330$
(EB), with $74{,}228$ ($96\%$) overlap. The higher threshold led to more
genes identified as DE in the MRF model. The numbers of DE genes
identified are presented in Figure~\ref{DEdataS}, where each line
represents a brain region. The table of the exact numbers is presented
in the supplementary material Section~9 [\citet{lin2015markovs}]. For
the number of DE genes, the trend over time was slightly different from
that in the previous section. In addition to the peak close to birth,
there was another peak that spanned from early childhood (period 10) to
adolescence (period 12). The peak was less obvious in the 5
nonneocortex regions (AMY, HIP, STR, MD and CBC). During these periods,
motor skills, social skills, emotional skills and cognitive skills are
rapidly developed. The second peak may correspond to the development of
these essential skills. Genes that were DE in the second peak may be of
interest to researchers studying these behaviors. Note that there was a
slight decrease in DE genes in periods 5--6 compared with that in
periods 4--5. The decrease was most obvious in brain region STR. Further
biological studies are needed to understand the trend.
We randomly split the data into two subsets and implemented the
algorithm separately for each subset. Compared with the EB model, the
genes identified as DE by the MRF model were more likely to overlap:
56.2\% vs. 12.4\% (supplementary material Section~9 [\citet{lin2015markovs}]). The information for the direction of changes in
gene expression was not utilized in the model. However, we observed
that DE genes in all neocortex regions tended to have the same
direction of changes (Table~\ref{ratioDE}). Therefore, the MRF model
is able to detect consistent changes in gene expression among the brain
regions, which may be missed by other approaches not considering
temporal and spatial similarity.

\begin{table}
\caption{Summary for the direction of changes in gene expression by
pooling neocortex regions. Each row represents a time window. The ``0''
column represents the counts of genes that were down-regulated in all
neocortex regions and the ``11'' column represents the counts of genes
that were up-regulated in all neocortex regions}
\label{ratioDE}
\begin{tabular*}{\tablewidth}{@{\extracolsep{\fill}}lcccccccccccc@{}}
\hline
& \textbf{0} & \textbf{1} & \textbf{2} & \textbf{3} & \textbf{4} &
\textbf{5} & \textbf{6} & \textbf{7} & \textbf{8} & \textbf{9} & \textbf
{10} & \textbf{11} \\
\hline
Periods 3--4   & \phantom{0}163 & \phantom{0}5 & 1 & 0 & 0 & 0 & 0 & 0 & 0 & 0 & \phantom{0}1  & \phantom{00}47   \\
Periods 4--5   & 1039           & 31           & 3 & 3 & 0 & 0 & 1 & 3 & 0 & 4 & 18 & \phantom{0}436  \\
Periods 5--6   & \phantom{0}539 & 30           & 3 & 1 & 1 & 0 & 1 & 0 & 0 & 2 & 20 & \phantom{0}417  \\
Periods 6--7   & 3475           & 28           & 3 & 2 & 1 & 1 & 2 & 2 & 0 & 2 & 29 & 1238 \\
Periods 7--8   & 1014           & 14           & 1 & 0 & 0 & 0 & 0 & 0 & 0 & 1 & \phantom{0}3  & 1640 \\
Periods 8--9   & \phantom{0}387 & \phantom{0}5            & 0 & 0 & 0 & 0 & 0 & 0 & 0 & 0 & \phantom{0}1  & \phantom{0}146  \\
Periods 9--10  & 1034           & \phantom{0}1            & 0 & 0 & 0 & 0 & 0 & 0 & 0 & 1 & \phantom{0}1  & \phantom{0}351  \\
Periods 10--11 & \phantom{0}342 & \phantom{0}2            & 0 & 0 & 0 & 0 & 0 & 0 & 0 & 0 & \phantom{0}3  & 1124 \\
Periods 11--12 & \phantom{0}915 & \phantom{0}9            & 0 & 0 & 0 & 0 & 0 & 0 & 0 & 0 & \phantom{0}1  & \phantom{0}485  \\
Periods 12--13 & \phantom{0}450 & \phantom{0}0            & 0 & 0 & 0 & 0 & 0 & 0 & 0 & 0 & \phantom{0}1  & \phantom{0}204  \\
Periods 13--14 & \phantom{0}263 & \phantom{0}5            & 0 & 0 & 0 & 0 & 0 & 0 & 0 & 0 & \phantom{0}2  & \phantom{00}39   \\
Periods 14--15 & \phantom{0}107 & 22           & 0 & 0 & 0 & 0 & 0 & 0 & 0 & 0 & \phantom{0}5  & \phantom{0}149  \\
\hline
\end{tabular*}
\end{table}

\begin{figure}

\includegraphics{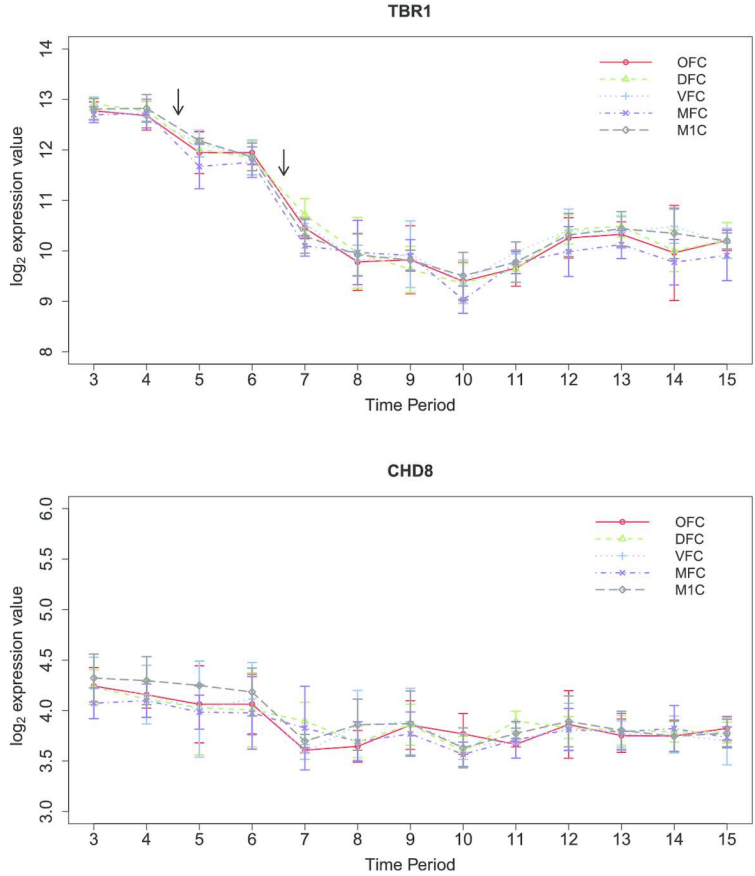}

\caption{The dynamics of gene expression for TBR1 and CHD8 in frontal
cortex regions. In periods 4--5 and 6--7, TBR1 was differentially
expressed in all frontal cortex regions, as indicated by the arrows in
the figure.}
\label{expvalue}\label{chd8}\label{tbr1}
\end{figure}

Autism Spectrum Disorders (ASD) are a group of syndromes characterized
by fundamental impairments in
social reciprocity and language development accompanied by highly
restrictive interests and/or repetitive behaviors [\citet{american2000diagnostic}]. By exome sequencing, loss of function (LoF)
mutations with large biological effects have been shown to affect ASD
risk [\citet{iossifov2012novo,kong2012rate,neale2012patterns}, \citeauthor{o2011exome} (\citeyear{o2011exome,o2012sporadic}),  \citet{sanders2012novo}].
A set of nine high-confidence ASD risk genes have been identified
recently: ANK2, CHD8, CUL3, DYRK1A, GRIN2B, KATNAL2, POGZ, SCN2A, TBR1
[\citet{willsey2013coexpression}]. These nine genes carry LoF mutations
in ASD patients. Details\vadjust{\goodbreak} for the genes are described in the
supplementary material Section~7 [\citet{lin2015markovs}]. Next we
analyzed the nine ASD risk genes in the human brain gene expression
data set. Among the nine genes, KATNAL2 and CHD8 were unexpressed. The
other seven genes were expressed in all brain regions and all periods.
Gene expression study on postmortem autistic brains and structural
magnetic resonance imaging studies have highlighted the frontal cortex
as pathological in ASD patients [\citet{amaral2008neuroanatomy,voineagu2011transcriptomic}]. In the brain gene
expression data, five regions were sampled in the frontal cortex: OFC,
DFC, VFC, MFC and M1C. The gene expression curves for TBR1 and CHD8 are
shown in Figure~\ref{expvalue}. The five frontal cortex regions shared
similar dynamics for the two genes. TBR1 was differentially expressed
in periods 4--5 and 6--7, while CHD8 remained unexpressed.
We performed a binomial test to see whether the ASD gene set was
enriched for DE genes, compared with the overall distribution
(Table~\ref{pvalueDE}). In the binomial test, a gene was counted as
DE only if it was DE in all five frontal cortex regions. We observed an
increased fold change of DE genes in the ASD gene set in periods 4--5,
5--6, 6--7, 9--10 and 10--11. It is interesting to note the gap that
spanned periods 7 to 9, when the ASD genes tended to be equally
expressed. For periods 4--5 and 9--10, the enrichment was significant
($<$0.05). Period 10 corresponds to early childhood ($1 \leq \mathrm{Age}\leq 6$), when social, emotional and cognitive skills are observed
[\citet{kang2011spatio}]. The most obvious signs of autism tend to emerge
between 2 and 3 years of age. In periods 9--10, there were four DE
genes: SCN2A, CUL3, ANK2, GRIN2B. These four genes are of potential
interest, as a malfunction of these genes in ASD patients may directly
affect the development of social and cognitive skills in early\vadjust{\goodbreak} childhood.

\begin{table}
\tablewidth=250pt
\caption{Enrichment analysis of DE genes in the ASD gene set}
\label{pvalueDE}
\begin{tabular*}{250pt}{@{\extracolsep{\fill}}lcccc@{}}
\hline
& \textbf{\# of DE} & \textbf{\# of DE} & \textbf{Fold change} & \textbf{$\bolds{p}$-value} \\
& \textbf{(expected)} &\textbf{(ASD)} & & \\
\hline
Periods 3--4 & 0.3 & 0 & 0\phantom{.0} & 0.62\phantom{0} \\
Periods 4--5 & 1.6 & 4 & 2.5 & 0.03\phantom{0} \\
Periods 5--6 & 1.2 & 3 & 2.5 & 0.06\phantom{0} \\
Periods 6--7 & 3.7 & 6 & 1.6 & 0.05\phantom{0} \\
Periods 7--8 & 2.1 & 0 & 0\phantom{.0} & 0.96\phantom{0} \\
Periods 8--9 & 0.4 & 0 & 0\phantom{.0} & 0.67\phantom{0} \\
Periods 9--10 & 1.0 & 4 & 3.9 & 0.006 \\
Periods 10--11 & 1.1 & 2 & 1.8 & 0.19\phantom{0} \\
Periods 11--12 & 1.1 & 1 & 0.9 & 0.50\phantom{0} \\
Periods 12--13 & 0.6 & 0 & 0\phantom{.0} & 0.72\phantom{0} \\
Periods 13--14 & 0.3 & 0 & 0\phantom{.0} & 0.64\phantom{0} \\
Periods 14--15 & 0.2 & 0 & 0\phantom{.0} & 0.60\phantom{0} \\
\hline
\end{tabular*}
\end{table}
%
\section{Simulation studies}\label{sec5}
\subsection{Identify expressed and unexpressed genes}\label{sec51}
We conducted simulation studies to evaluate the performance of our
proposed MRF model. The expression values for 100 genes in 16 brain
regions and 13 periods were simulated. The number of replicates was set
to be 3. The latent state array was first simulated and we considered
two simulation settings:

\subsubsection*{Simulation setting 1} The latent state array was simulated by
Gibbs sampling. The sampler started from a random array with equal
probability of being expressed or unexpressed. The latent states were
updated sequentially by \eqref{cp} and the MRF parameters were set to
$\gamma=0.08$, $\beta_{1}=0.20$ and $\beta_{2}=1.5$. We conducted three
rounds of Gibbs sampling to obtain the latent state array $\mathbf{X}$.

\subsubsection*{Simulation setting 2} In period 1, all genes had equal
probability of being unexpressed/expressed. The latent states evolved
over time by a Hidden Markov Model with 0.1 transition probability. The
latent states for the 16 brain regions were initially set to be the
same. Then we let different proportions $(0.1, 0.2, 0.5)$ of the latent
states flip randomly.

The gene expression levels were simulated based on the latent states.
The mean gene expression array $\bolds{\mu}$ was generated from $\mathbf{X}$
by a Gaussian mixture model, where $\mu_1=4.5$, $\sigma_1=0.75$, $\mu
_2=(5,5.5,6,6.5,7,7.5,8)$ and $\sigma_2=1.5$. We varied $\mu_2$ and
kept the other parameters unchanged to test the model in different
scenarios. Parameters were set to be the same for all brain regions.
The gene expression levels $\mathbf{Y}$ were then simulated from a normal
distribution, with mean $\bolds{\mu}$ and variance $\sigma^2_0=0.25$. The
MCEM algorithm and the Gibbs sampler were implemented the same as in
the previous sections. A comparison of misclassification rates was made
between the MRF model and the simple Gaussian mixture model with no
temporal and spatial dependency assumed (Table~\ref{misclass}). For all
simulation settings, the MRF model achieved significant improvement in
misclassification rates compared with the simple Gaussian mixture model.

\begin{table}
\caption{Comparison of misclassification rates between the simple
Gaussian mixture model (GMM) and the MRF model. The standard deviations
in 100 independent runs are shown in the brackets. The results for
simulation settings 1 and 2 are presented, the numbers after the model
names represent the proportions $(0.1, 0.2, 0.5)$ of purturbation in
simulation setting 2}
\label{misclass}
\begin{tabular*}{\tablewidth}{@{\extracolsep{\fill}}lcccc@{}}
\hline
$\bolds{\mu_2}$ & \textbf{GMM} & \textbf{MRF} & \textbf{GMM (0.1)} & \textbf
{MRF (0.1)}\\
\hline
5 & 0.421 (0.025) & 0.093 (0.008) & 0.426 (0.012) & 0.131 (0.004) \\
5.5 & 0.346 (0.017) & 0.084 (0.006) & 0.375 (0.013) & 0.11 (0.004)\phantom{0}\\
6 & 0.275 (0.011) & 0.071 (0.005) & 0.31 (0.014)\phantom{0} & 0.093 (0.002) \\
6.5 & 0.203 (0.006) & 0.055 (0.004) & 0.242 (0.009) & 0.083 (0.002)\\
7 & 0.144 (0.004) & 0.041 (0.003) & 0.185 (0.006) & 0.072 (0.003)\\
7.5 & 0.101 (0.003) & 0.029 (0.002) & 0.137 (0.004) & 0.053 (0.002)\\
8 & 0.067 (0.002) & 0.020 (0.001) & 0.096 (0.004) & 0.037 (0.002)\\[6pt]
$\bolds{\mu_2}$ & \textbf{GMM (0.2)} & \textbf{MRF (0.2)} & \textbf
{GMM (0.5)} & \textbf{MRF (0.5)} \\
\hline
5 & 0.423 (0.008) & 0.233 (0.005) & 0.421 (0.004) & 0.344 (0.008) \\
5.5 & 0.378 (0.011) & 0.208 (0.005) & 0.377 (0.005) & 0.312 (0.011) \\
6 & 0.31 (0.012)\phantom{0} & 0.18 (0.004)\phantom{0} & 0.309 (0.004) & 0.261 (0.007)\\
6.5 & 0.242 (0.009) & 0.144 (0.004) & 0.243 (0.004) & 0.187 (0.004)\\
7 & 0.185 (0.004) & 0.106 (0.003) & 0.185 (0.004) & 0.133 (0.003)\\
7.5 & 0.137 (0.004) & 0.075 (0.002) & 0.138 (0.003) & 0.093 (0.002)\\
8 & 0.096 (0.003) & 0.051 (0.002) & 0.096 (0.003) & 0.060 (0.002) \\
\hline
\end{tabular*}
\end{table}
%
\subsection{Identify DE genes over time}\label{sec52}
In the simulation study, data were generated for 100 genes, 16 brain
regions and 12 periods. We considered three simulation settings:

\subsubsection*{Simulation setting 1}
The latent state array $\mathbf{S}$ was
updated sequentially by~\eqref{postDE} and the MRF parameters were set
to $\gamma_{\mathrm{DE}}=-0.10$, $\beta_{\mathrm{cc}}=0.31$, $\beta_{\mathrm{nn}}=0.52$, $\beta
_{\mathrm{cn}}=0.06$ and $\beta_{t}=0.14$. To keep the ratio of DE genes roughly
the same as that in the real data, the sampler started from a random
array with 0.4 probability of being DE. $10\%$ of the genes were then
randomly selected to be unexpressed in all brain regions from periods
$1$ to $t$ or $t$ to $T=12$, where $t$ was randomly picked from
$1,\ldots,T$. The presence of unexpressed genes reflects the fact that a
small portion of genes switched their states of unexpressed/expressed
in the real data. We conducted three rounds of Gibbs sampling to obtain
the latent state array $\mathbf{S}$. The $z$-score array $\mathbf{Z}$ was then
generated from $\mathbf{S}$ by a mixture model. For EE, the $z$-score was
generated from $\mathcal{N}(0,1)$; for DE, it was generated from
$\mathcal{N}(-2,1)$ or $\mathcal{N}(2,1)$, with equal probability.

\subsubsection*{Simulation setting 2}
The latent state array $\mathbf{S}$ was
simulated by Gibbs sampling with the same setting as in \textit{simulation setting $1$}. The mean gene expression array $\bolds{\mu}$ was
then generated from $\mathbf{S}$. In period 1, all the genes had mean
expression values at 0. From period $t$ to $t+1$, $\mu_{bg(t+1)}=\mu
_{bgt}+s_{bgt}\delta$, where $\delta\sim\mathcal{N}(0,1)$. Finally, the
gene expression\vspace*{1pt} array $\mathbf{Y}$ was generated from $\bolds{\mu}$ by
Gaussian distribution with variance $\sigma_0^2=0.25$ and the number of
replicates was set to be 3.

\subsubsection*{Simulation setting 3} In period 1, all the genes had 0.15
probability of being DE. From periods $t$ to $t+1$, 70\% of the DE
genes in period $t$ randomly switched to EE, and the same number of EE
genes randomly switched to DE, to keep the number of DE genes constant
over time. To represent the neocortex and nonneocortex regions, the
first 11 brain regions were set to have the same latent states and the
other 5 brain regions were set to be the same. Compared with the first
11 brain regions, 40\% of the DE genes randomly switched to EE in the
other 5 brain regions. Then we randomly selected different proportions
$(0.1, 0.2, 0.5)$ of the DE states to switch to EE; the same number of EE
states were randomly selected to switch to DE. 10\% of the genes were
randomly selected to be unexpressed in all brain regions as in \textit{simulation setting $1$}. Finally, the $z$-score array $\mathbf{Z}$ was
generated in the same way as in \textit{simulation setting $1$}.

\begin{figure}
\begin{tabular}{@{}cc@{}}

\includegraphics{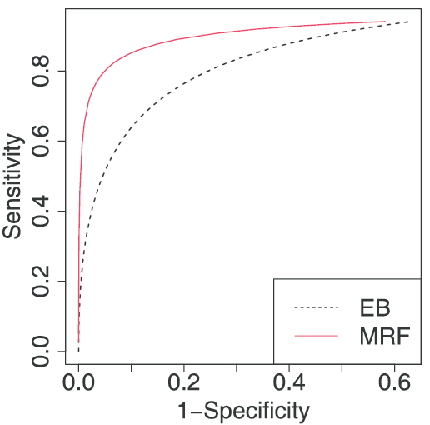}
 & \includegraphics{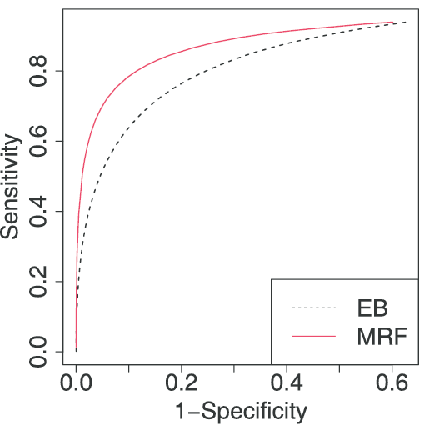}\\
\footnotesize{(a) Setting 1, neocortex} & \footnotesize{(b) Setting 1, nonneocortex}\\[3pt]

\includegraphics{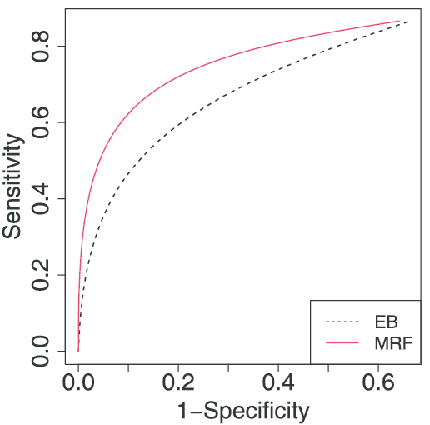}
 & \includegraphics{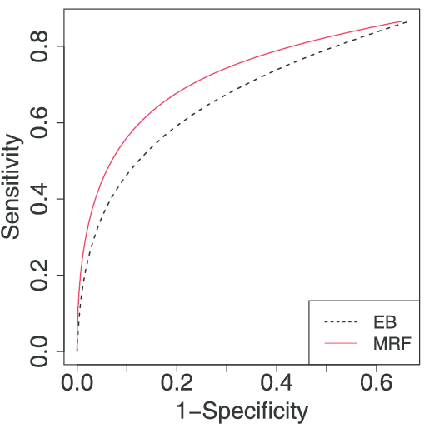}\\
\footnotesize{(c) Setting 2, neocortex}& \footnotesize{(d) Setting 2, nonneocortex}\\[3pt]

\includegraphics{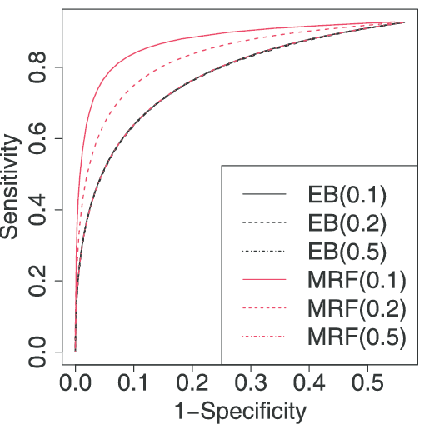}
 & \includegraphics{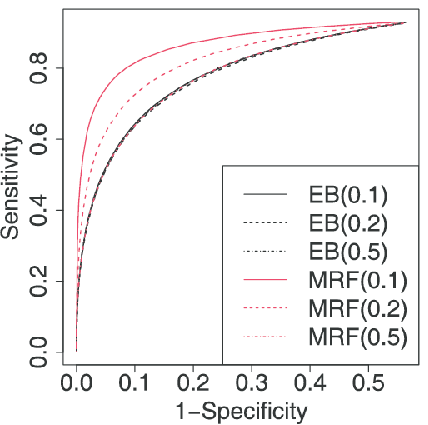}\\
\footnotesize{(e) Setting 3, neocortex}& \footnotesize{(f) Setting 3, nonneocortex}
\end{tabular}
\caption{The ROC curves comparing the empirical Bayesian (EB) model and
the proposed MRF model. The curves were averaged over 100 simulations.}
\label{roc}\label{sc1}\label{sn1}\label{sc2}\label{sn2}\label{sc3}\label{sn3}
\end{figure}

The settings for the MCEM algorithm and the Gibbs sampler were the same
as those in the previous section. We calculated the sensitivity and
specificity by varying the threshold for the posterior local-f.d.r. We
compared the proposed MRF model with the empirical Bayesian (EB) model,
which assumes no temporal and spatial dependency (Figure~\ref{roc}). As
the neocortex group and the nonneocortex group have different numbers
of brain regions (11 vs. 5), the ROC curves were plotted separately for
the two groups. Compared with the EB model, the MRF model performed
better in both the neocortex and nonneocortex regions. The improvement
was more significant in the neocortex regions, as there were more brain
regions and the MRF model benefits more from the spatial similarity.
%
%
\section{Conclusions and discussion}\label{sec6}
The statistical methods developed in this paper were motivated from the
analysis of human brain development microarray data. These data
represent expression profiles in different brain regions at different
developmental stages and they allow us to infer (1) whether a gene is
expressed or not in a specific brain region in a specific period, and
(2) whether a gene is differentially expressed between two adjacent
periods in a specific brain region. To efficiently utilize the spatial
similarity between brain regions and temporal dependency, we have
developed a two-step modeling framework that is based on the Markov
Random Field model and local FDR methodology to facilitate statistical
inference. Our simulation studies suggest that this model has a lower
misclassification rate compared with commonly used Gaussian mixture
models without considering spatial similarity and temporal dependency.
Simulation results and real data analysis also suggest that the
proposed model improves the power to identify DE genes.

The analysis of the human brain microarray data by our proposed model
produces biologically meaningful results. The inferred latent states of
``expressed'' or ``unexpressed'' were similar in all brain regions. The
number of genes that switched their latent states first increased and
peaked at birth, then gradually decreased in adulthood. In periods 6--7,
the list of genes that switched from expressed to unexpressed was
enriched for transcriptional regulatory genes. For the purpose of
identifying DE genes between adjacent periods, we observed a similar
trend in the number of DE genes. However, there was an additional peak
in periods that correspond to childhood and adolescence. These
observations reflect the dynamics of the neurodevelopment process. We
also observed that genes carrying a high risk for neurodevelopment
disorders, such as ASD, tended to be differentially expressed,
especially during periods when cognitive and social skills were developed.

We have also proposed and implemented an MCEM algorithm to estimate the
model parameters and a separate Gibbs sampler to estimate the posterior
probability. In previous studies, the iterated conditional mode (ICM)
algorithm was implemented to estimate the MRF parameters [\citet
{wei2007statistical,li2010hidden,besag1986statistical}]; however, our
simulation study suggested that the ICM algorithm may lead to biased
parameter estimates (supplementary material Section~10 [\citet
{lin2015markovs}]). One limitation of the MCEM algorithm is the high
computing cost. Under the current setting for the MCEM algorithm, the
computing time for the whole data set took ten days (five days for
biological question 1 and five days for biological question 2) on the
Yale Louise high performance cluster (Dell m620 system, 8 core
processor, 48 GB of memory). To accelerate convergence, we started the
model from the estimation which does not consider the spatial and
temporal dependency. Another limitation of the MCEM algorithm is that
the Monte Carlo sum is an approximation to the expectation and may lead
to instability in parameter estimation. In the diagnosis of the MCEM
algorithm (supplementary material Section~5 [\citet{lin2015markovs}]),
we demonstrated that our model is robust to unstable parameter
estimation. \citet{levine2001implementations} provided a detailed
discussion on the setting of the MCEM algorithm.

\begin{appendix}
\section*{Appendix}\label{app}
We provide details on the derivation of the conditional probability
\eqref{cp} from the joint probability \eqref{jointunexp}.

For $t \neq1 $ and $t \neq T $,
%
\begin{eqnarray*}
\nonumber
&& \frac{p(x_{bgt}=1 \vert \mathbf{X}/x_{bgt};\bolds{\Phi})}{p(x_{bgt}=0 \vert
\mathbf{X}/x_{bgt};\bolds{\Phi})} \\
&&\qquad = \frac{p(x_{bgt}=1,\mathbf{X}/x_{bgt};\bolds{\Phi})}{p(x_{bgt}=0,
\mathbf{X}/x_{bgt};\bolds{\Phi})}
\\
&&\qquad =  \exp \biggl\{\gamma_{1}-\gamma_{0}+
\beta_{1} \sum_{b'\neq
b} \bigl[I_{1}(x_{b'gt})-I_{0}(x_{b'gt})
\bigr]
\\
&&\hspace*{9pt}\qquad\qquad {}+ \beta _{2} \bigl[I_{1}(x_{bg(t-1)})-I_{0}(x_{bg(t-1)})+I_{1}(x_{bg(t+1)})-I_{0}(x_{bg(t+1)})
\bigr] \biggr\}
\\
&&\qquad = \exp\biggl\{ \gamma+ \beta_{1}\sum_{b'\neq
b}(2x_{b'gt}-1)+
\beta_{2}[2x_{bg(t-1)}-1+2x_{bg(t+1)}-1]\biggr\},
\end{eqnarray*}
$p(x_{bgt}=1 \vert \mathbf{X}/x_{bgt};\bolds{\Phi}) + p(x_{bgt}=0 \vert
\mathbf{X}/x_{bgt};\bolds{\Phi}) = 1$, so we have
\[
p(x_{bgt}=1\vert \mathbf{X}/x_{bgt};\bolds{\Phi})=
\frac{\exp\{F(x_{bgt},\bolds{\Phi})\}}{1+\exp\{F(x_{bgt},\bolds{\Phi})\}},
\]
where
\[
F(x_{bgt},\bolds{\Phi})=\gamma + \beta_{1}\sum
_{b'\neq
b}(2x_{b'gt}-1)+\beta_{2}
\{2x_{bg(t-1)}-1+2x_{bg(t+1)}-1\}.
\]
For $t=1 $ and $t=T$, the conditional probability can be derived similarly.
\end{appendix}

\section*{Acknowledgments}
We thank Christopher Fragoso for useful comments and suggestions on the
manuscript. We also thank the three anonymous reviewers, the anonymous
Associate Editor and the Area Editor Karen Kafadar for the
conscientious efforts and helpful comments. The analysis in this
article was performed at the Yale University Biomedical High
Performance Computing Center.

Conflict of interest: none declared.

\begin{supplement}[id=suppA]
\stitle{Supplement to ``A Markov random field-based approach to
characterizing human brain development using spatial--temporal
transcriptome data''}
\slink[doi]{10.1214/14-AOAS802SUPP} 
\sdatatype{.pdf}
\sfilename{aoas802\_supp.pdf}
\sdescription{Section~1: More information on the brain regions. Section~2: Spatial
and temporal similarity. Section~3: Microarray quality control
procedures. Section~4: Model fit and the robustness of the Gaussian
mixture model. Section~5: Diagnosis for the MCEM algorithm. Section~6:
Gene Ontology (GO) enrichment analysis. Section~7: High confidence ASD
genes. Section~8: Supplementary data for Section~\ref{sec41}. Section~9:
Supplementary data for Section~\ref{sec42}. Section~10: Comparison between the
ICM algorithm and the MCEM algorithm.}
\end{supplement}




\printaddresses
\end{document}